\documentclass[reprint,twocolumn,superscriptaddress,amsmath,amssymb,aps,prb,]{revtex4-2}
\usepackage{graphicx}
\usepackage{bm}
\usepackage{subcaption}  
\usepackage{lmodern}        % 添加现代字体支持
\usepackage{mathtools}      % 增强的数学公式支持
\usepackage[colorlinks=true, linkcolor=blue, urlcolor=blue, citecolor=blue, pdfusetitle]{hyperref}

\begin{document}

\title{Near-flat-band-driven violation of Pauli limit \\in heavy fermion superconductors}

\author{Yan-Xiao Wang}
\affiliation{Key Laboratory of Quantum Theory and Applications of MoE \& School of Physical Science and Technology, Lanzhou University, Lanzhou 730000, China}
\affiliation{Lanzhou Center for Theoretical Physics, Key Laboratory of Theoretical Physics of Gansu Province, Lanzhou University, Lanzhou 730000, China}

\author{Yin Zhong}
\email{zhongy@lzu.edu.cn}
\affiliation{Key Laboratory of Quantum Theory and Applications of MoE \& School of Physical Science and Technology, Lanzhou University, Lanzhou 730000, China}
\affiliation{Lanzhou Center for Theoretical Physics, Key Laboratory of Theoretical Physics of Gansu Province, Lanzhou University, Lanzhou 730000, China}

\begin{abstract}
Heavy-fermion superconductors often display upper critical fields that exceed the conventional Pauli paramagnetic limit, indicating that strong correlations and hybridized quasiparticle bands play an essential role in the paramagnetic pair-breaking process. Within the two-dimensional Kondo--Heisenberg model, we perform a self-consistent mean-field analysis of spin-singlet $s$-, extended-$s$-, and $d$-wave pairing under Zeeman fields, and compute the critical field $B_c$, transition temperature $T_c$, and Clogston--Chandrasekhar ratio $r_{\mathrm{CC}}$. We find that $r_{\mathrm{CC}}$ increases sharply as the conduction filling approaches half filling. This enhancement arises from the weakly dispersive region of the lower hybridized band, where the strongly reduced Fermi velocity diminishes the normal-state paramagnetic energy and stabilizes superconductivity. At fixed filling, the distinct $J_H$ dependences among the three pairing channels reflect the sensitivity of Pauli limiting to both band curvature and the structure of the order parameter. These results provide microscopic evidence that proximity to a near-flat hybridized band offers a robust route to enhanced Pauli-limiting fields in heavy-fermion superconductors.
\end{abstract}

\maketitle

\section{Introduction}

Heavy-fermion superconductors provide an ideal setting for exploring the interplay between strong electronic correlations, Kondo hybridization, and unconventional pairing mechanisms. A striking aspect of several Ce-based compounds is the highly nontrivial response of their superconducting states to magnetic fields. CeCoIn$_5$ is a paradigmatic Pauli-limited spin-singlet superconductor, and its low-temperature, high-field behavior exhibits features consistent with a Fulde--Ferrell--Larkin--Ovchinnikov (FFLO) state~\cite{Bianchi2003FFLO,ThompsonJPSJ2012}. In sharp contrast, UTe$_2$ displays extraordinarily large and highly anisotropic upper critical fields that far exceed the singlet Pauli limit, consistent with proposed triplet or multicomponent order parameters~\cite{Ran2019Science,Hayes2021Science}. More recent experiments have further expanded this landscape: high-pressure CeSb$_2$ exhibits a pronounced enhancement of $H_{c2}$ beyond the conventional Pauli limit~\cite{PhysRevLett.131.026001}, while the two-phase superconductor CeRh$_2$As$_2$ shows unusually large and strongly anisotropic upper critical fields in both superconducting phases~\cite{Onishi2022CeRh2As2}. Together with recent theoretical studies on the Clogston--Chandrasekhar limit and its possible violation~\cite{PhysRevLett.9.266,PhysRevB.108.214511,PhysRevB.107.174516,PhysRevB.105.L060501}, these results highlight a central unresolved question: under what circumstances can a predominantly spin-singlet system substantially exceed the conventional Pauli paramagnetic limit? These diverse high-field behaviors raise the need for a microscopic description that captures both heavy-fermion quasiparticles and their sensitivity to magnetic fields.

To address this issue, it is essential to employ microscopic frameworks in which heavy quasiparticles and their hybridized band structures emerge self-consistently. The Kondo model describes the antiferromagnetic exchange between conduction electrons and localized $f$ moments, generating heavy quasiparticles with reconstructed Fermi surfaces in the coherent regime~\cite{Si2001,PhysRevLett.99.136401,PhysRevLett.101.066404}. The Kondo--Heisenberg (KH) model extends this description by incorporating a local-moment Heisenberg interaction that captures short-range Ruderman--Kittel--Kasuya--Yosida (RKKY) correlations~\cite{PhysRevB.56.11820,PhysRevB.69.035111,PhysRevB.72.245111,PhysRevLett.98.026402}. Within large-$N$ fermionic formulations, the KH model naturally produces hybridization, spin-liquid correlations, and several competing spin-singlet pairing instabilities---including $s$-, extended-$s$-, and nodal or nodeless $d$-wave states---whose relative stability depends sensitively on filling, $J_H/J_K$, and the resulting Fermi-surface geometry~\cite{PhysRevLett.57.877,PhysRevB.35.3394,PhysRevLett.62.595,Liu2014CPL,sato2001strong,PhysRevLett.100.146403}. This makes the KH model a minimal platform for studying how heavy-band formation influences superconducting properties. Given this microscopic structure, it is crucial to understand how different singlet pairing channels within the KH model respond to magnetic fields.

Because Pauli limiting is highly sensitive to both the quasiparticle dispersion and the structure of the superconducting gap, a framework that captures multiple heavy-fermion pairing channels is required. Experiments indeed reveal a rich variety of spin-singlet states: nodal $d$-wave pairing in CeCu$_2$Si$_2$ and CeCoIn$_5$~\cite{stockert2011magnetically,PhysRevLett.87.057002,aoki2004field,allan2013imaging,zhou2013visualizing}, and fully gapped $s$-wave behavior in CeRu$_2$ and CeCo$_2$~\cite{matsuda1995101ru,PhysRevLett.94.057001}. Several Ce-based compounds also exhibit strong Pauli-limited behavior—and in some cases clear exceedance of the Clogston--Chandrasekhar estimate—underscoring the close connection between superconductivity and heavy hybridized quasiparticle bands~\cite{PhysRevB.106.184509}. These considerations motivate a unified microscopic study of Pauli limiting across multiple competing singlet pairing channels within the KH model. These experimental trends also resonate with broader developments in multiband and weak-dispersion superconductivity.

Recent progress in multiband and flat-band superconductivity has demonstrated that weakly dispersive bands and strong-coupling effects can greatly enhance spin-singlet pairing and, under suitable conditions, produce effective violations of the Pauli limit~\cite{PhysRevLett.130.226001,Cao2021}. However, such approaches typically introduce pairing interactions phenomenologically and do not generate hybridization or pairing self-consistently from microscopic Kondo or Heisenberg exchange couplings. Consequently, it remains unclear whether these mechanisms can naturally operate in heavy-fermion systems, where heavy quasiparticles originate directly from Kondo hybridization. This motivates a systematic analysis of Pauli limiting within a minimal Kondo framework. This motivates a controlled microscopic study in which hybridization, correlations, and superconducting pairing all emerge self-consistently.

In this work, we revisit the Pauli limit of spin-singlet superconductivity in the two-dimensional KH model by computing the combined thermal and Zeeman suppression of pairing within a fully self-consistent mean-field formalism. Building on the established zero-field phase diagram~\cite{Liu2014CPL}, we determine the field--temperature phase boundaries $T_c(B)$ for several pairing channels, extract the zero-temperature critical field $B_c$, and evaluate the corresponding Pauli-limit ratio
\begin{equation}
r_{\mathrm{CC}}=\frac{\mu_B B_c}{1.25 k_B T_c}.
\end{equation}
By systematically varying the conduction-band filling, local exchange couplings, and pairing symmetry, we identify regimes in which $r_{\mathrm{CC}}$ is substantially enhanced and show that this enhancement originates from the proximity of the Fermi level to weakly dispersive regions of the hybridized $cf$ band. Our results demonstrate that even within a purely spin-singlet framework with an isotropic $g$ factor, strong correlations and Kondo hybridization can significantly elevate the Pauli limit through a flat-band mechanism, providing a controlled theoretical baseline for interpreting the anomalously large upper critical fields observed in several heavy-fermion superconductors.

The remainder of this paper is organized as follows. Sec.~\ref{sec:level2} introduces the KH model in a Zeeman field and our self-consistent mean-field approach. Sec.~\ref{sec:level3} presents the resulting field--temperature phase diagrams and Pauli-limit ratios. Sec.~\ref{sec:level4} provides a flat-band interpretation of the enhanced Pauli limit. Sec.~\ref{sec:level5} summarizes the main conclusions.

\section{Model and Method}\label{sec:level2}

We begin by introducing the Kondo–Heisenberg model, whose Hamiltonian is given in Ref.~\cite{Liu2014CPL}
\begin{equation}
H=\sum_{\boldsymbol{k},\sigma}\epsilon_{\boldsymbol{k}}\,
c_{\boldsymbol{k}\sigma}^\dagger c_{\boldsymbol{k}\sigma}
+J_K\sum_i \mathbf{S}_i\cdot \mathbf{s}_i
+J_H\sum_{\langle i,j\rangle}\mathbf{S}_i\cdot \mathbf{S}_j .
\end{equation}
The localized f-electron moments are represented by the Abrikosov fermions
\begin{equation}
\mathbf{S}_i=\frac12 f_{i\alpha}^\dagger\boldsymbol{\tau}_{\alpha\beta}f_{i\beta},
\qquad  
\sum_{\sigma} f_{i\sigma}^\dagger f_{i\sigma}=1 ,
\end{equation}
where the $f$ fermions (spinons) carry only the spin degrees of freedom. Within the large-$N$ fermionic formulation, the Kondo and Heisenberg exchanges can be rewritten as
\begin{align}
\mathbf{S}_i\cdot \mathbf{s}_i
&=-\frac12 (f_{i\uparrow}^\dagger c_{i\uparrow}+f_{i\downarrow}^\dagger c_{i\downarrow})
(c_{i\uparrow}^\dagger f_{i\uparrow}+c_{i\downarrow}^\dagger f_{i\downarrow}), \\
\mathbf{S}_i\cdot\mathbf{S}_j
&=-\frac12 (f_{i\uparrow}^\dagger f_{j\uparrow}+f_{i\downarrow}^\dagger f_{j\downarrow})
(f_{j\uparrow}^\dagger f_{i\uparrow}+f_{j\downarrow}^\dagger f_{i\downarrow}),
\end{align}
where constant terms have been omitted.  These expressions naturally motivate the introduction of the uniform hybridization and resonating-valence-bond (RVB) order parameters
\begin{equation}
V=-\sum_\sigma \langle c_{i\sigma}^\dagger f_{i\sigma}\rangle ,\qquad
\chi=-\sum_\sigma \langle f_{i\sigma}^\dagger f_{j\sigma}\rangle .
\end{equation}

To avoid accidental degeneracies in the conduction band on the square lattice, we adopt the dispersion
\begin{equation}
\epsilon_{\boldsymbol{k}}=-2t(\cos k_x+\cos k_y)
+4t'\cos k_x\cos k_y-\mu ,
\end{equation}
where the chemical potential $\mu$ is tuned to fix the conduction electron density $n_c$.  Under the uniform mean-field approximation, the spinons acquire a narrow dispersion
\begin{equation}
\chi_{\boldsymbol{k}} = J_H \chi (\cos k_x+\cos k_y)+\lambda ,
\end{equation}
with the Lagrange multiplier $\lambda$ enforcing the single-occupancy constraint on average.

The resulting hybridized normal-state Hamiltonian is
\begin{equation}
\mathcal{H}
=\sum_{\boldsymbol{k},\sigma}
\begin{pmatrix}
c_{\boldsymbol{k}\sigma}^\dagger & f_{\boldsymbol{k}\sigma}^\dagger
\end{pmatrix}
\begin{pmatrix}
\epsilon_{\boldsymbol{k}} & \tfrac{J_K}{2}V \\
\tfrac{J_K}{2}V & \chi_{\boldsymbol{k}}
\end{pmatrix}
\begin{pmatrix}
c_{\boldsymbol{k}\sigma}\\ f_{\boldsymbol{k}\sigma}
\end{pmatrix}
+E_0 ,
\end{equation}
with
\begin{equation}
E_0 = N_s\left(
J_H \chi^2
+ \frac12 J_K V^2
- \lambda
+ \mu n_c
\right).
\end{equation}
Diagonalization yields two hybridized quasiparticle bands
\begin{equation}\label{1}
E_{\boldsymbol{k}}^{(\pm)} =\frac12\left[\epsilon_{\boldsymbol{k}}+\chi_{\boldsymbol{k}}\pm\sqrt{\left(\epsilon_{\boldsymbol{k}}-\chi_{\boldsymbol{k}}\right)^2+ (J_K V)^2}\right].
\end{equation}

% ----------------------------------------------------------------------

Having established the hybridized normal state, we next incorporate an external magnetic field. The Zeeman coupling is
\begin{equation}
H_Z = -\frac{\mu_B B}{2}\sum_{j,\sigma}\sigma
\left(
g_c\, c_{j\sigma}^\dagger c_{j\sigma}
+ g_f\, f_{j\sigma}^\dagger f_{j\sigma}
\right).
\end{equation}
Throughout this work we set $\mu_B=k_B=1$ and $g_c=g_f=2$, which reduces the Zeeman term to
\begin{equation}
H_Z = - B \sum_{j,\sigma}\sigma
\left(
c_{j\sigma}^\dagger c_{j\sigma}
+ f_{j\sigma}^\dagger f_{j\sigma}
\right).
\end{equation}

Beyond hybridization, the antiferromagnetic (AFM) Heisenberg exchange also supports spinon pairing. Using the fermionic representation, it can be written as
\begin{equation}
\mathbf{S}_i\cdot \mathbf{S}_j
= -\frac12
\big(f_{i\uparrow}^\dagger f_{j\downarrow}^\dagger
      - f_{i\downarrow}^\dagger f_{j\uparrow}^\dagger\big)
\big(f_{j\downarrow} f_{i\uparrow}
      - f_{j\uparrow} f_{i\downarrow}\big).
\end{equation}
As emphasized by Coleman \textit{et al.}, the SU(2) gauge structure of the local moments requires both hopping and pairing channels. We introduce the spinon pairing order parameter
\begin{equation}
\Delta_{ij}
= -\left\langle
f_{i\uparrow}^\dagger f_{j\downarrow}^\dagger
- f_{i\downarrow}^\dagger f_{j\uparrow}^\dagger
\right\rangle .
\end{equation}
On the square lattice, $\Delta_{ij}$ carries either extended-$s$ or $d$-wave symmetry.  In momentum space this becomes $\Delta_f\gamma_{\boldsymbol{k}}$, where
\begin{equation}
\gamma_{\boldsymbol{k}} =
\begin{cases}
\cos k_x - \cos k_y, & d_{x^2-y^2}\text{-wave},\\
\cos k_x + \cos k_y, & \text{extended }s\text{-wave}.
\end{cases}
\end{equation}

The Kondo exchange likewise contains an on-site singlet pairing channel
\begin{equation}
\mathbf{S}_i\cdot \mathbf{s}_i
= -\frac12
\big(c_{i\uparrow}^\dagger f_{i\downarrow}^\dagger
      - c_{i\downarrow}^\dagger f_{i\uparrow}^\dagger\big)
\big(f_{i\downarrow} c_{i\uparrow}
      - f_{i\uparrow} c_{i\downarrow}\big),
\end{equation}
leading to the local $s$-wave conduction--spinon pairing order parameter
\begin{equation}
\Delta_{cf}
= -\left\langle
c_{i\uparrow}^\dagger f_{i\downarrow}^\dagger
- c_{i\downarrow}^\dagger f_{i\uparrow}^\dagger
\right\rangle .
\end{equation}
The hybridization $V$ and RVB hopping $\chi$ retain their definitions from the normal-state analysis.

To treat the pairing states, we introduce the Nambu spinor
\begin{equation}
\Psi_{\boldsymbol{k}} =
\begin{pmatrix}
c_{\boldsymbol{k}\uparrow}\\ f_{\boldsymbol{k}\uparrow}\\
c_{-\boldsymbol{k}\downarrow}^\dagger\\ f_{-\boldsymbol{k}\downarrow}^\dagger
\end{pmatrix},
\end{equation}
which leads to the Bogoliubov--de Gennes Hamiltonian
\begin{equation}
\mathcal H(\boldsymbol{k})=
\begin{pmatrix}
\epsilon_{\boldsymbol{k}} - B & \tfrac{J_K}{2}V & 0 & \tfrac{J_K}{2}\Delta_{cf}\\[3pt]
\tfrac{J_K}{2}V & \chi_{\boldsymbol{k}} - B & \tfrac{J_K}{2}\Delta_{cf} & J_H\Delta_f\gamma_{\boldsymbol{k}}\\[3pt]
0 & \tfrac{J_K}{2}\Delta_{cf} & -\epsilon_{\boldsymbol{k}} - B & -\tfrac{J_K}{2}V\\[3pt]
\tfrac{J_K}{2}\Delta_{cf} & J_H\Delta_f\gamma_{\boldsymbol{k}} & -\tfrac{J_K}{2}V & -\chi_{\boldsymbol{k}} - B
\end{pmatrix}.
\end{equation}
Its four quasiparticle branches are
\begin{align}
E_{\boldsymbol{k}1} &= E_{\boldsymbol{k}H} - B, \\
E_{\boldsymbol{k}2} &= E_{\boldsymbol{k}L} - B, \\
E_{\boldsymbol{k}3} &= -E_{\boldsymbol{k}H} - B, \\
E_{\boldsymbol{k}4} &= -E_{\boldsymbol{k}L} - B ,
\end{align}
where the positive-energy eigenvalues at $B=0$ are
\begin{equation}
E_{\boldsymbol{k}H} = \frac{1}{2}\sqrt{E_{\boldsymbol{k}} + 2\sqrt{E_{\boldsymbol{k}0}}}, 
\qquad
E_{\boldsymbol{k}L} = \frac{1}{2}\sqrt{E_{\boldsymbol{k}} - 2\sqrt{E_{\boldsymbol{k}0}}},
\end{equation}
with
\begin{align}
E_{\boldsymbol{k}} &= 
J_K^2\left(\Delta_{cf}^2 + V^2\right)
+ 2J_H^2\Delta_f^2\gamma_{\boldsymbol{k}}^2
+ 2\epsilon_{\boldsymbol{k}}^2
+ 2\chi_{\boldsymbol{k}}^2 , \\
E_{\boldsymbol{k}0} &= 
\big[J_K\Delta_{cf}(\epsilon_{\boldsymbol{k}} - \chi_{\boldsymbol{k}})
     - 2J_H\Delta_f\gamma_{\boldsymbol{k}}\, \epsilon_{\boldsymbol{k}} V \big]^2 \nonumber \\
&\quad
+ J_H^2\Delta_f^2\gamma_{\boldsymbol{k}}^2
  \big(J_K^2\Delta_{cf}^2
      + J_K^2 V^2
      + 2\chi_{\boldsymbol{k}}^2
      - 2\epsilon_{\boldsymbol{k}}^2
      + J_H^2\Delta_f^2\gamma_{\boldsymbol{k}}^2 \big)
      \nonumber \\
&\quad
+ J_K^2 V^2 (\epsilon_{\boldsymbol{k}} + \chi_{\boldsymbol{k}})^2
+ \left(\epsilon_{\boldsymbol{k}}^2 - \chi_{\boldsymbol{k}}^2\right)^2 .
\end{align}

At temperature $T$, the free energy is  
\begin{align}
F &= -T\sum_{\boldsymbol{k}}\sum_{\alpha=H,L}\Big[\ln(1+e^{-\beta(E_{\boldsymbol{k}\alpha}-B)})+\ln(1+e^{-\beta(E_{\boldsymbol{k}\alpha}+B)})\Big]  \nonumber\\
&\quad +\sum_{\boldsymbol{k}}(\epsilon_{\boldsymbol{k}}+\chi_{\boldsymbol{k}}-E_{\boldsymbol{k}H}-E_{\boldsymbol{k}L})+ E_0 ,
\end{align}
with
\begin{equation}
E_0 = N_s\left[\frac{J_K}{2}(V^2+\Delta_{cf}^2)+J_H(\chi^2+\Delta_f^2)-\lambda+\mu n_c\right].
\end{equation}

Variation of $F$ with respect to  
\[
V,\ \Delta_{cf},\ \chi,\ \Delta_f,\ \lambda,\ \mu
\]
yields the saddle-point equations.  

To quantify how close the Fermi level lies to the weakly dispersive region of the lower hybridized band, we evaluate the following normal-state quantities for each set of parameters:
\begin{itemize}

\item the energy separation between the Fermi level and the top of the lower
hybridized band
\[
E_1^{\rm top}-E_F;
\]

\item the Fermi velocity of the lower hybridized band~\cite{ashcroft_solid_1976}
\[
v_F(\boldsymbol{k}) = \bigl|\nabla_{\boldsymbol{k}}E_1(\boldsymbol{k})\bigr|,
\]
and its Fermi-surface average
\[
\langle v_F \rangle_{\rm FS}
=\frac{1}{L_{\rm FS}}
\oint_{\rm FS} dk\, v_F(k),
\]
together with its inverse $1/\langle v_F \rangle_{\rm FS}$ as a measure of the band flatness;

\item the density of states at the Fermi level in two dimensions~\cite{ashcroft_solid_1976,Yanase2003PhysRep}
\[
N(0)=\frac{1}{2\pi^{2}}
\oint_{\rm FS}\frac{dk}{v_F(k)},
\]
where the expression already includes the twofold spin degeneracy;

\item the uniform normal-state spin susceptibility~\cite{ashcroft_solid_1976,Kittel2005}
\[
\chi_n = \left.\frac{\partial M}{\partial B}\right|_{B\to 0},
\]
where $M(B)$ is the magnetization obtained from the Zeeman splitting
\[
E_{\boldsymbol k\uparrow}=E_{\boldsymbol k}-\mu_B B, \qquad E_{\boldsymbol k\downarrow}=E_{\boldsymbol k}+\mu_B B,
\]
and the difference between the spin-resolved occupations is evaluated at a low temperature.
\end{itemize}

These quantities characterize how increasing $n_c$ drives the system toward the weakly dispersive (near-flat) regime of the hybridized band, and how varying $J_H$ tunes the system toward or away from this regime. They provide the microscopic foundation for understanding the evolution of the Clogston--Chandrasekhar ratio discussed in the main text.

\section{Results}\label{sec:level3}

Unless otherwise specified, we set \(t'/t=0.3\) and \(J_K/t=2.0\) throughout the following analysis.

\subsection{Ground-state phase diagram of the superconducting state}

\begin{figure}
  \centering
  \includegraphics[width=0.45\textwidth]{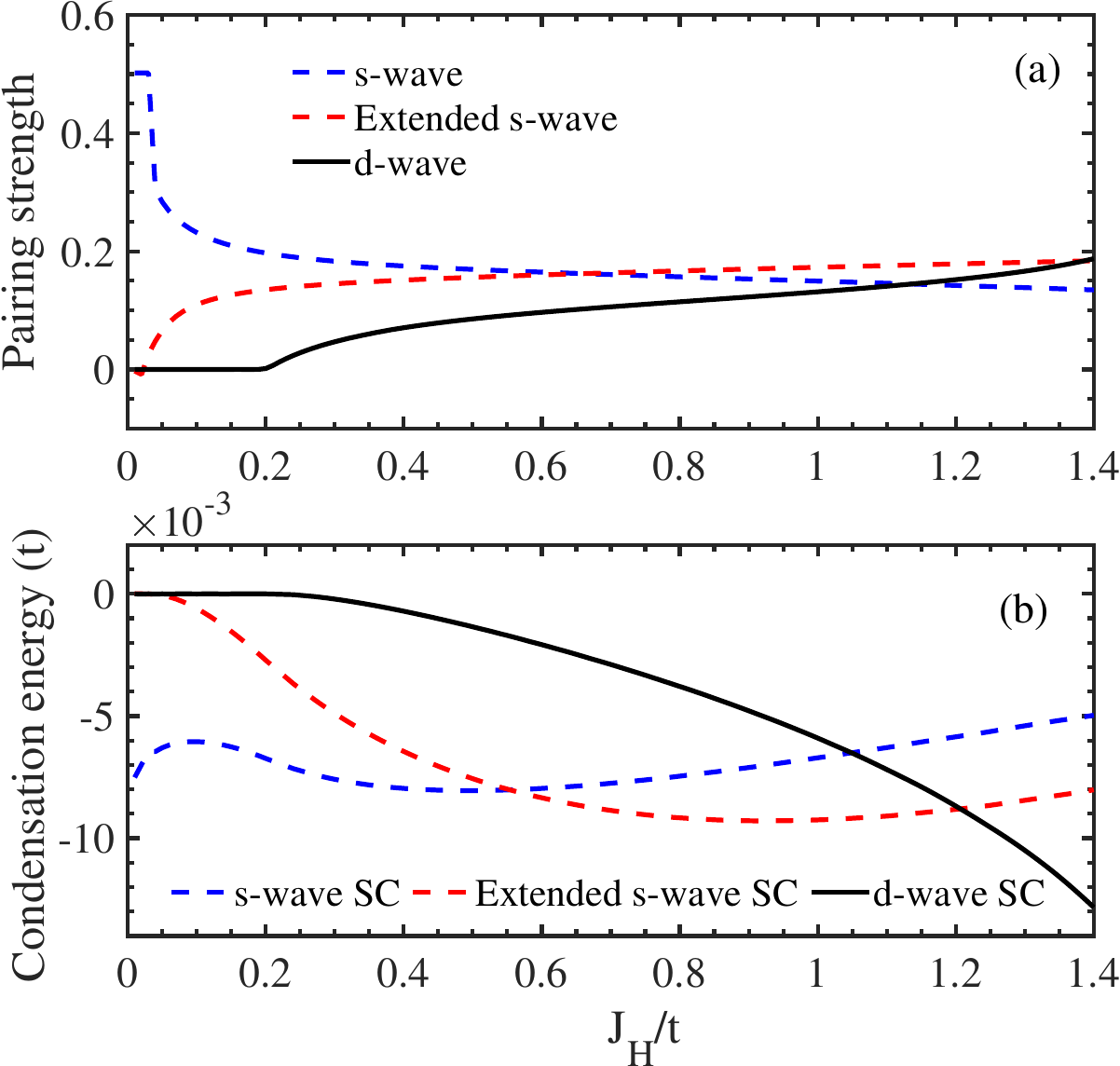}
  \caption{(Color online) (a) Evolution of the pairing amplitudes \(\Delta_{cf}\) and \(\Delta_{f}\) as a function of the local AFM exchange \(J_H/t\).
  (b) Condensation energy \(E_{\rm cond}=\Omega_{\rm SC}-\Omega_{\rm N}\) (per site, in units of \(t\)) for the \(s\)-, extended-\(s\)-, and \(d\)-wave states.
  Model parameters: \(t'/t=0.3\), \(n_c=0.91\), and \(J_K/t=2\).
  Line styles and colors follow the legend.}
  \label{fig2}
\end{figure}

\begin{figure}
  \centering
  \includegraphics[width=0.40\textwidth]{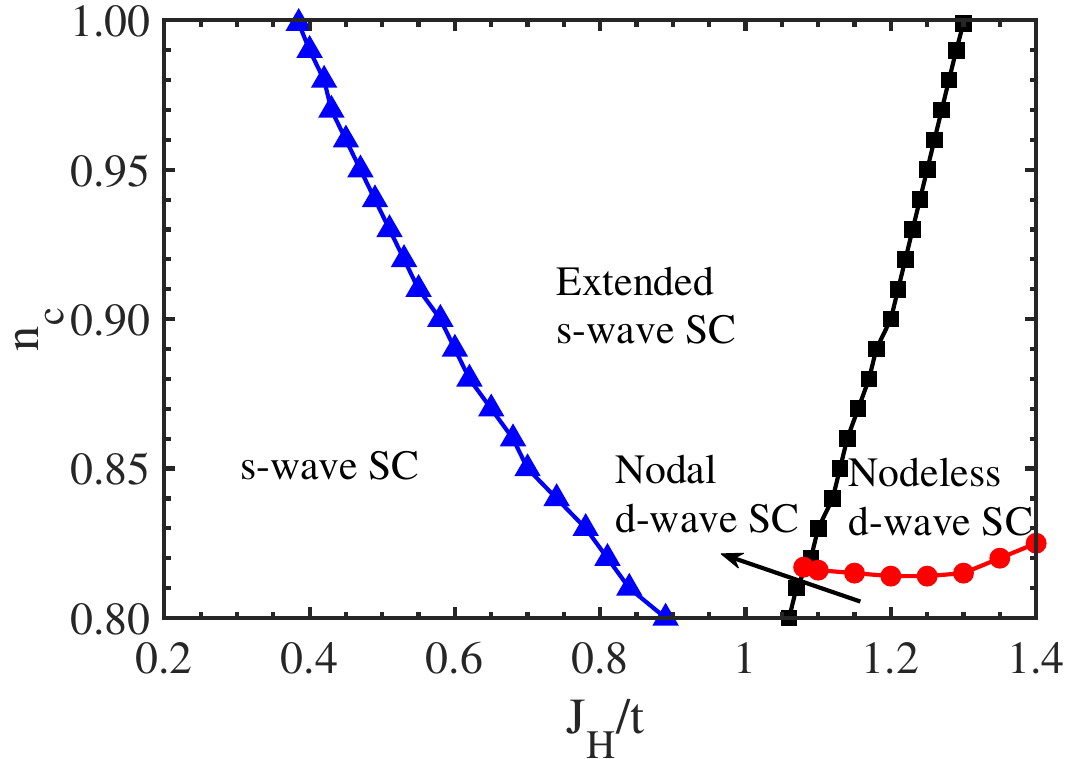}
  \caption{(Color online) Zero-field superconducting ground-state phase diagram for $t'/t=0.3$ and $J_K/t=2.0$. The red curve marks the boundary between nodal and 
nodeless $d$-wave states, obtained from our self-consistent solutions following the standard criterion in Ref.~\cite{Liu2014CPL}.}
  \label{fig3}
\end{figure}

We first tested ``coexisting'' solutions with simultaneous electron--spinon and spinon--spinon pairing (\(\Delta_{cf}\Delta_f\neq0\)) and found them unstable within the mean-field iterations. We therefore focus on three stable channels: \(s\)-wave (\(\Delta_f=0\)), extended \(s\)-wave, and \(d\)-wave (\(\Delta_{cf}=0\)).
For \(t'/t=0.3\), \(n_c=0.91\), and \(J_K/t=2\), the self-consistent solutions are summarized in Fig.~\ref{fig2}.
As shown in Fig.~\ref{fig2}(a), \(\Delta_{cf}\) decreases while \(\Delta_f\) increases with \(J_H/t\); the \(d\)-wave channel sets in near \(J_H/t\simeq0.20\) and then grows approximately linearly.
The condensation-energy comparison in Fig.~\ref{fig2}(b) yields three regimes:
\(s\)-wave for \(0<J_H/t\lesssim0.55\), extended \(s\)-wave for \(0.55\lesssim J_H/t\lesssim1.21\), and \(d\)-wave for \(J_H/t\gtrsim1.21\).
Combining the amplitude evolution and the energy ordering gives the zero-field ground-state phase diagram in Fig.~\ref{fig3}, where the red curve separates nodal and nodeless $d$-wave states.

To study the field response we choose representative zero-field points deep inside each phase:
away from half filling, \((J_H/t,n_c)=(0.35,0.84)\) for \(s\)-wave, \((1.00,0.84)\) for extended \(s\), and \((1.35,0.81)\) for nodal \(d\)-wave;
near half filling, \((0.35,0.995)\), \((1.00,0.995)\), and \((1.35,0.995)\) for \(s\)-, extended-\(s\)-, and nodeless \(d\)-wave, respectively.
All chosen points lie well inside their phases, ensuring stable convergence of the self-consistent solutions.

\begin{figure}
  \centering
  \includegraphics[width=0.45\textwidth]{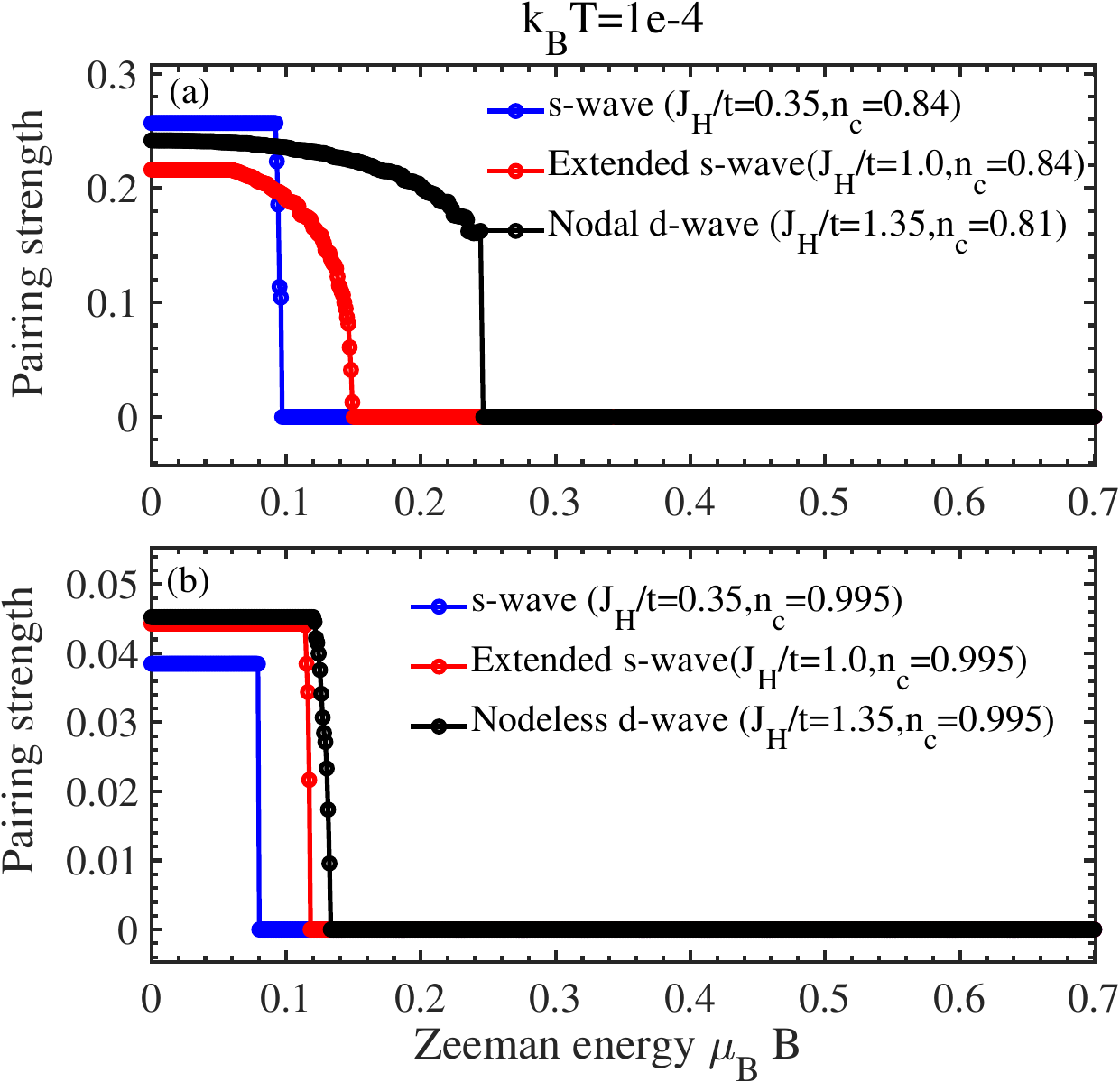}
  \caption{(Color online) Field dependence of the pairing amplitudes \(\Delta_{cf}\) and \(\Delta_f\) at \(k_{\mathrm B}T=10^{-4}\) for representative \(s\)-, extended-\(s\)-, and \(d\)-wave states.  Panels (a) and (b) correspond to moderately doped and near–half-filled cases, respectively. The pairing amplitudes collapse at a critical field \(B_c\), signaling a quasi–first-order suppression of superconductivity.}
  \label{fig4}
\end{figure}

\begin{figure}
  \centering
  \includegraphics[width=0.45\textwidth]{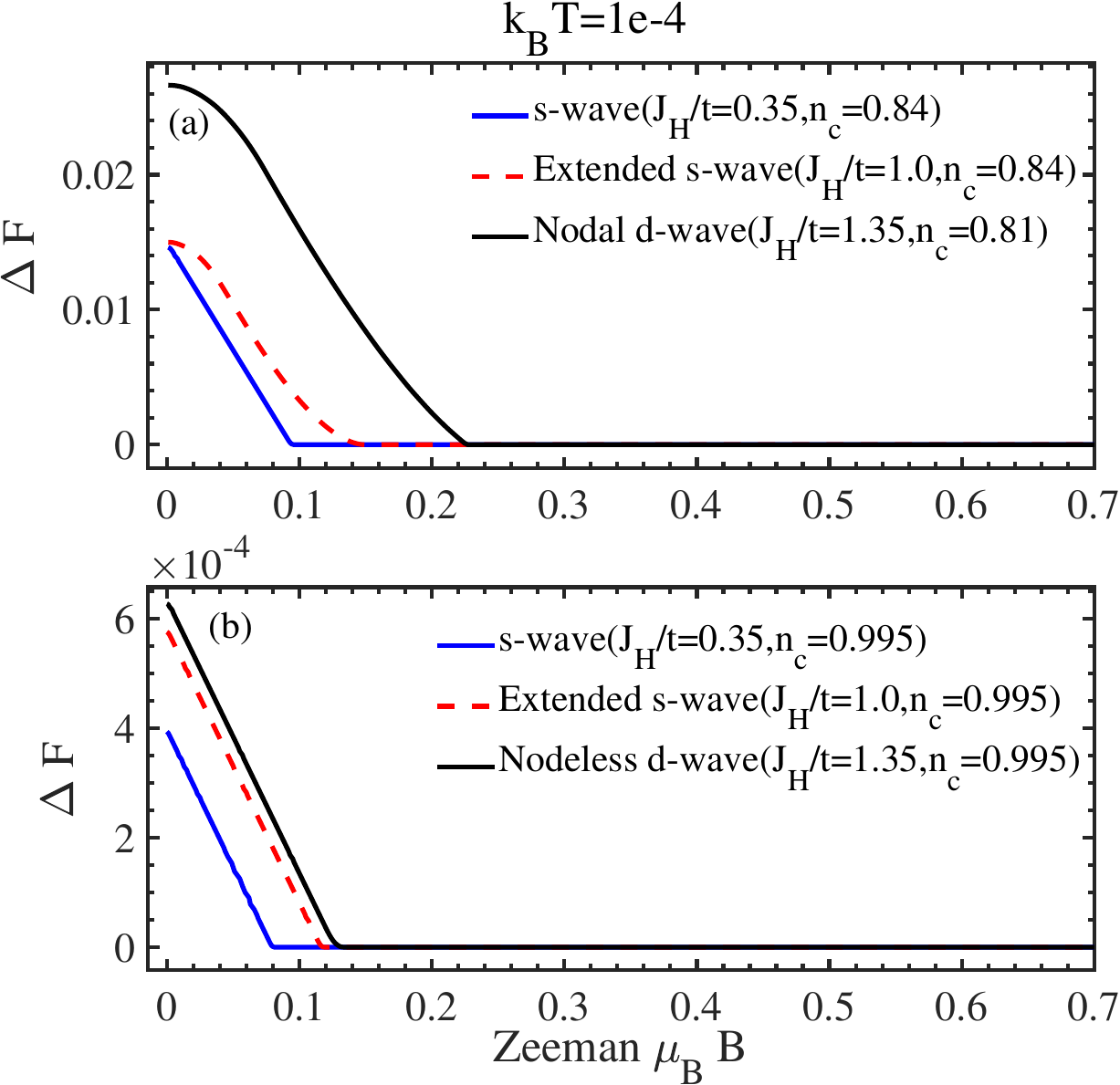}
  \caption{(Color online) Free-energy difference \(\Delta F = F_{\mathrm N} - F_{\mathrm{SC}}\) versus \(\mu_{\mathrm B}B\) at \(k_{\mathrm B}T=10^{-4}\).  
  Positive values favor superconductivity.}
  \label{fig5}
\end{figure}

\begin{figure}
  \centering
  \includegraphics[width=0.45\textwidth]{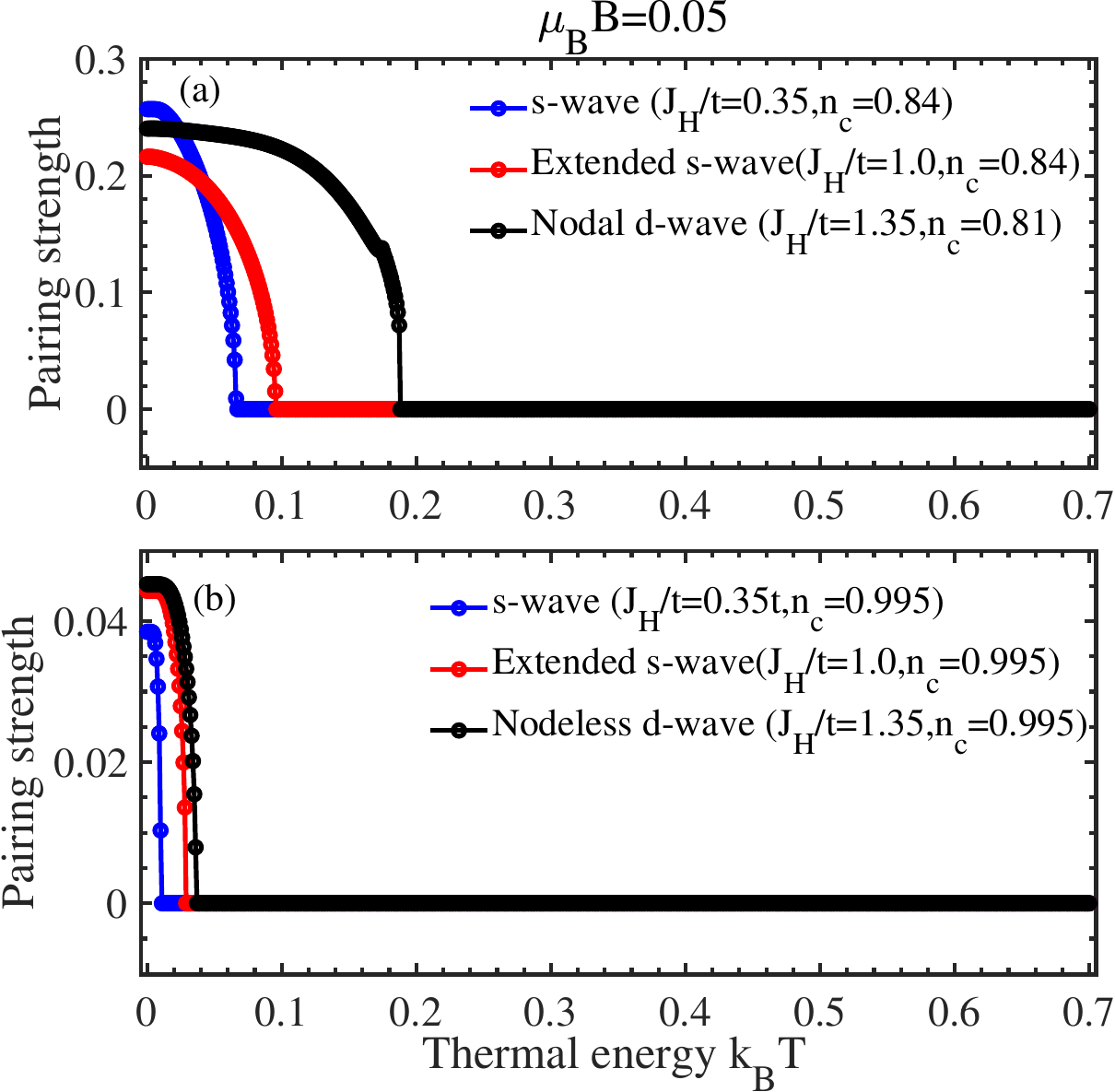}
  \caption{(Color online) Temperature dependence of \(\Delta_{cf}\) and \(\Delta_f\) at \(\mu_{\mathrm B}B=0.05\) for \(s\)-, extended-\(s\)-, 
  and \(d\)-wave states.  
  Panels (a) and (b) show moderately doped and near–half-filled cases, respectively. In each case the order parameters decrease smoothly with temperature and vanish at the corresponding $T_c(B)$.}
  \label{fig6}
\end{figure}

\subsection{Zeeman-field and temperature dependence of the pairing amplitudes}

At a fixed low temperature \(k_{\mathrm B}T = 10^{-4}\), we first examine the Zeeman response of representative parameter points selected from the zero-field phase diagram.  As shown in Fig.~\ref{fig4}, the pairing amplitudes for the \(s\)-, extended-\(s\)-, and \(d\)-wave states decrease monotonically with the Zeeman energy and collapse at a critical field \(B_c\), indicating a quasi--first-order suppression of superconductivity.  The free-energy difference in Fig.~\ref{fig5} yields the same \(B_c\), confirming the consistency of the two criteria.

To construct the full field--temperature phase boundary, we compute the temperature evolution of the pairing amplitudes at various fixed fields.  As an illustrative example, Fig.~\ref{fig6} shows the behavior at \(\mu_{\mathrm B}B=0.05\), where the order parameters decrease smoothly with temperature and vanish at the corresponding \(T_c\).  Repeating this procedure for different fields provides the complete \(T_c(B)\) dependence, which forms the basis for our discussion of the Clogston--Chandrasekhar limit.

\subsection{Field--temperature phase boundary and systematic trends of Pauli-limit enhancement}

\begin{figure}
  \centering
  \includegraphics[width=0.45\textwidth]{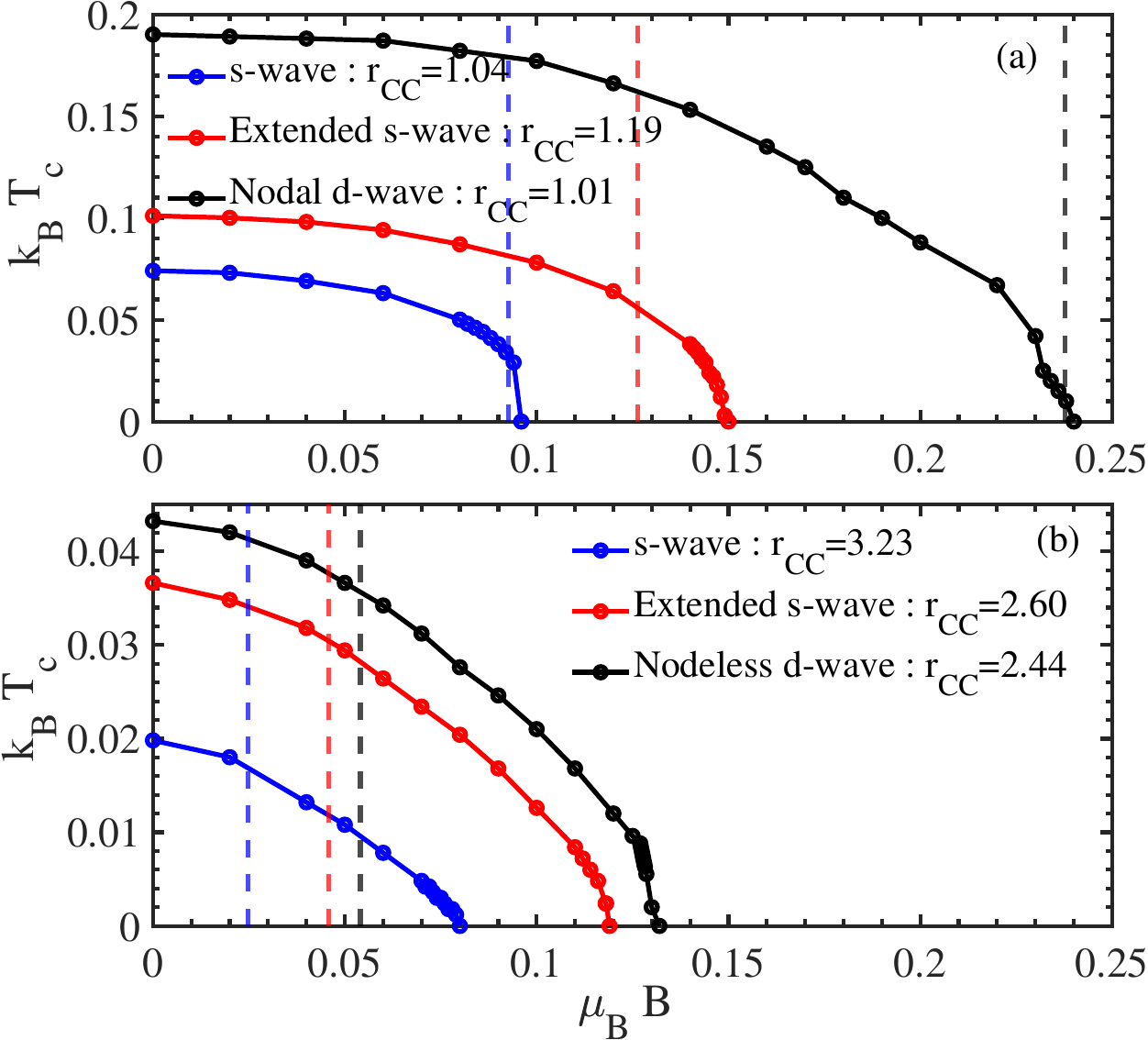}
  \caption{(Color online) (a–b) Field dependence of the superconducting transition temperature \(k_{\mathrm B}T_c(B)\) versus Zeeman field \(\mu_{\mathrm B}B\) for the \(s\)-, extended-\(s\)-, and \(d\)-wave states. \(T_c\) is obtained from \(\Delta_{cf(f)}(T\!\to\!T_c)\to0\); vertical dashed lines mark the Clogston--Chandrasekhar estimate \(\mu_{\mathrm B}B_{\mathrm{CC}}=1.25\,k_{\mathrm B}T_c(0)\). Legends report \(r_{\mathrm{CC}}\equiv \mu_{\mathrm B}B_c/[1.25\,k_{\mathrm B}T_c(0)]\). For moderate doping [panel (a)] all channels yield \(r_{\mathrm{CC}}\approx1\), while near half filling [panel (b)] the critical fields clearly exceed the Clogston--Chandrasekhar estimate. All other parameters \((J_H/t,\,n_c)\) are the same as in the preceding figure.}
  \label{fig7}
\end{figure}

\begin{figure}
  \centering
  \includegraphics[width=0.45\textwidth]{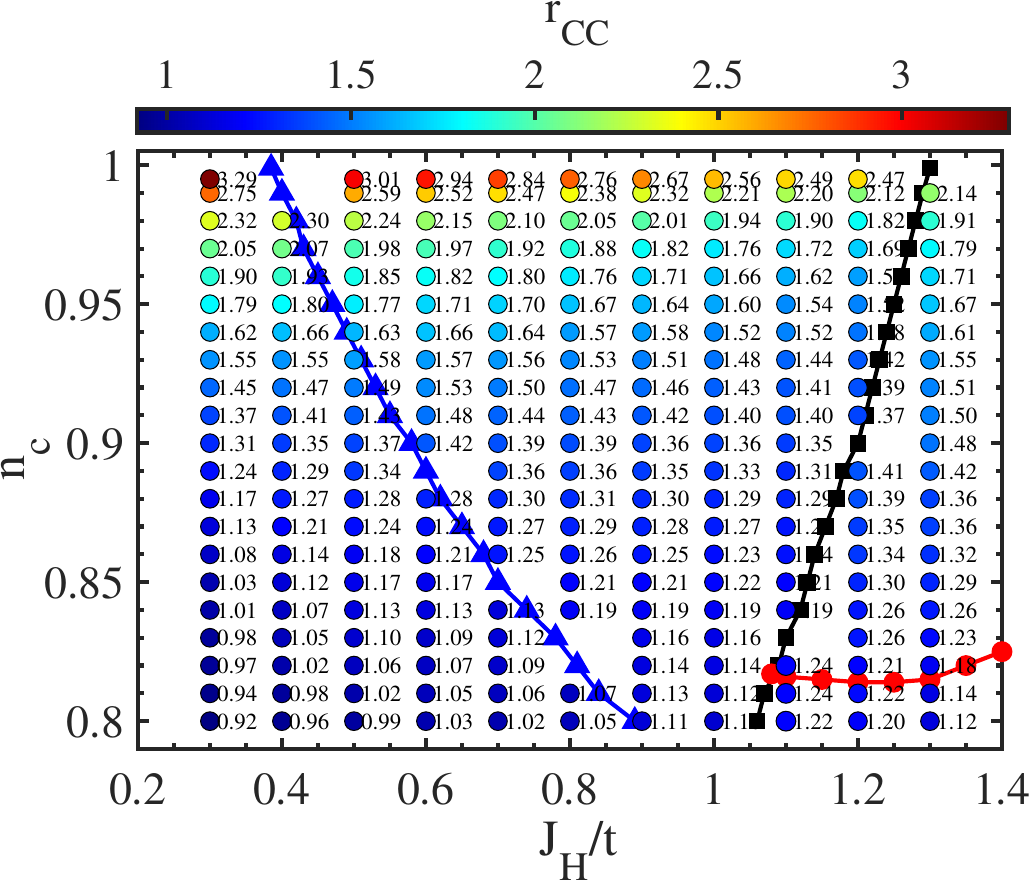}
  \caption{(Color online) Distribution of the Clogston--Chandrasekhar ratio  $r_{\mathrm{CC}}$ across the zero-field superconducting phase diagram of Fig.~\ref{fig3}. Colored circles denote the values of $r_{\mathrm{CC}}$. For each fixed $J_H/t$, $r_{\mathrm{CC}}$ increases monotonically with the filling and reaches its maximum near half filling ($n_c\!\to\!1$).}
  \label{fig8}
\end{figure}

Fig.~\ref{fig7} summarizes the critical lines \(T_c(B)\) obtained from fixed-field temperature scans.  Formally, the upper critical field \(B_c\) is
defined by the condition \(T_c(B\!\to\!B_c)=0\).  In practice, we extract \(B_c\) from the collapse of the self-consistent pairing amplitudes at \(k_{\mathrm B}T=10^{-4}\), which provides an accurate zero-temperature approximation.  For all parameter sets examined, this procedure yields values of \(B_c\) that agree with the extrapolation of \(T_c(B)\) within our numerical precision.  We therefore use the zero-temperature pairing-collapse criterion throughout the remainder of the analysis.

To quantify deviations from the Clogston--Chandrasekhar estimate, we introduce the dimensionless ratio
\[
r_{\mathrm{CC}}
   = \mu_{\mathrm B}B_c / [1.25\,k_{\mathrm B}T_c(0)] .
\]
For the moderately doped case shown in Fig.~\ref{fig7}(a), all three pairing channels yield \(r_{\mathrm{CC}}\approx1\), indicating no substantial violation
of the Pauli limit.  Near half filling [Fig.~\ref{fig7}(b)], however, \(r_{\mathrm{CC}}\) exceeds unity by a large margin, and the extracted \(B_c\)
lies well above the dashed Clogston--Chandrasekhar reference line.

To reveal the global evolution of Pauli limiting across the phase diagram, Fig.~\ref{fig8} maps \(r_{\mathrm{CC}}\) over a wide range of \((J_H/t,\,n_c)\).  A clear systematic trend emerges: for each fixed \(J_H/t\), \(r_{\mathrm{CC}}\) increases monotonically with the conduction-band filling and reaches its maximum as \(n_c\to1\).  This steady enhancement indicates that Pauli pair breaking becomes progressively weaker as the Fermi level approaches a weakly dispersive (near-flat) region of the hybridized band.  A microscopic flat-band interpretation of this behavior is developed in the next section.

\section{Flat-band mechanism for exceeding the conventional Pauli limit}\label{sec:level4}

\begin{figure}
  \centering
  \includegraphics[width=0.4\textwidth]{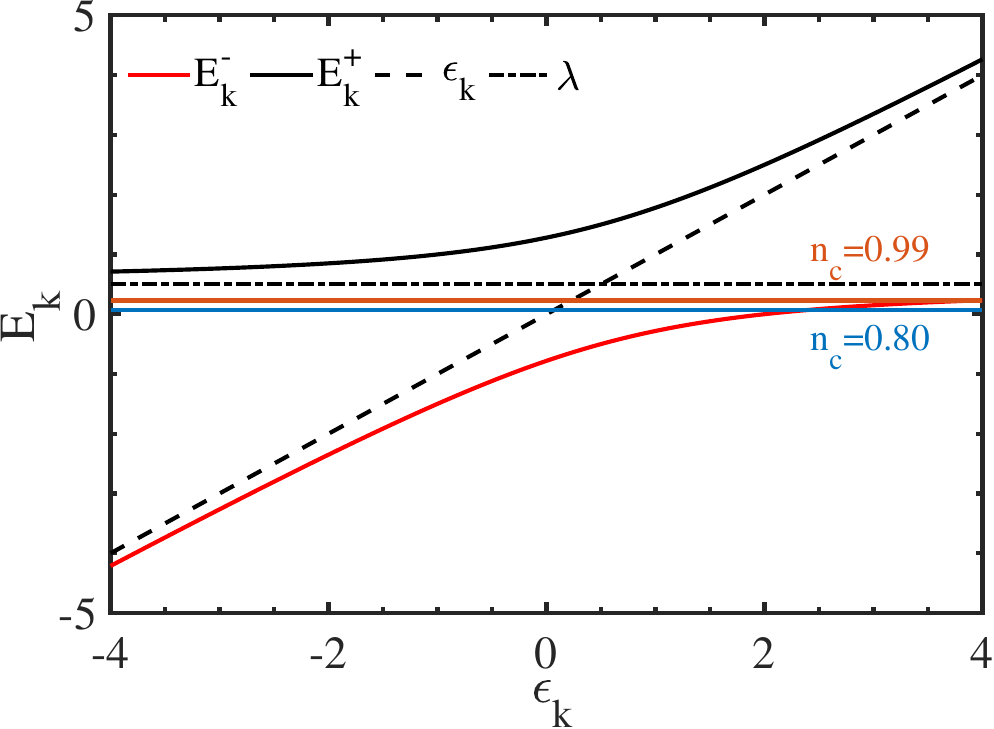}
  \caption{Schematic hybridized band structure.  The dispersive conduction band $\epsilon_k$ (black dashed) hybridizes with a nearly flat $f$ level $\lambda$ (black dash–dotted), producing the upper ($E_k^{+}$, black solid) and lower ($E_k^{-}$, red solid) hybridized bands.  Horizontal lines indicate representative Fermi levels for two fillings ($n_c=0.80$ and $n_c=0.99$), illustrating how increasing $n_c$ brings the Fermi level closer to the weakly dispersive region at the top of the lower band.}
  \label{fig11}
\end{figure}

%\begin{figure}
%  \centering
%  \includegraphics[width=0.3\textwidth]{fig/9FS(3).pdf}
%  \caption{Momentum-resolved dispersion of the lower hybridized band $E_{1}(\boldsymbol{k})$ at filling $n_{c}=0.995,J_H/t=1.0$. The color map shows the band energy, and the white dashed contour denotes the Fermi surface defined by $E_{1}=0$. The region enclosed by the green dashed curves marks the weakly dispersive portion of the band top, where $E_{1}(\boldsymbol{k})$ lies within a narrow window below the maximum $E_{1}^{\mathrm{top}}$, demonstrating that the top of the lower hybridized band forms a finite-area, nearly flat plateau.}
%  \label{fig14}
%\end{figure}

\begin{figure}
  \centering
  \includegraphics[width=0.5\textwidth]{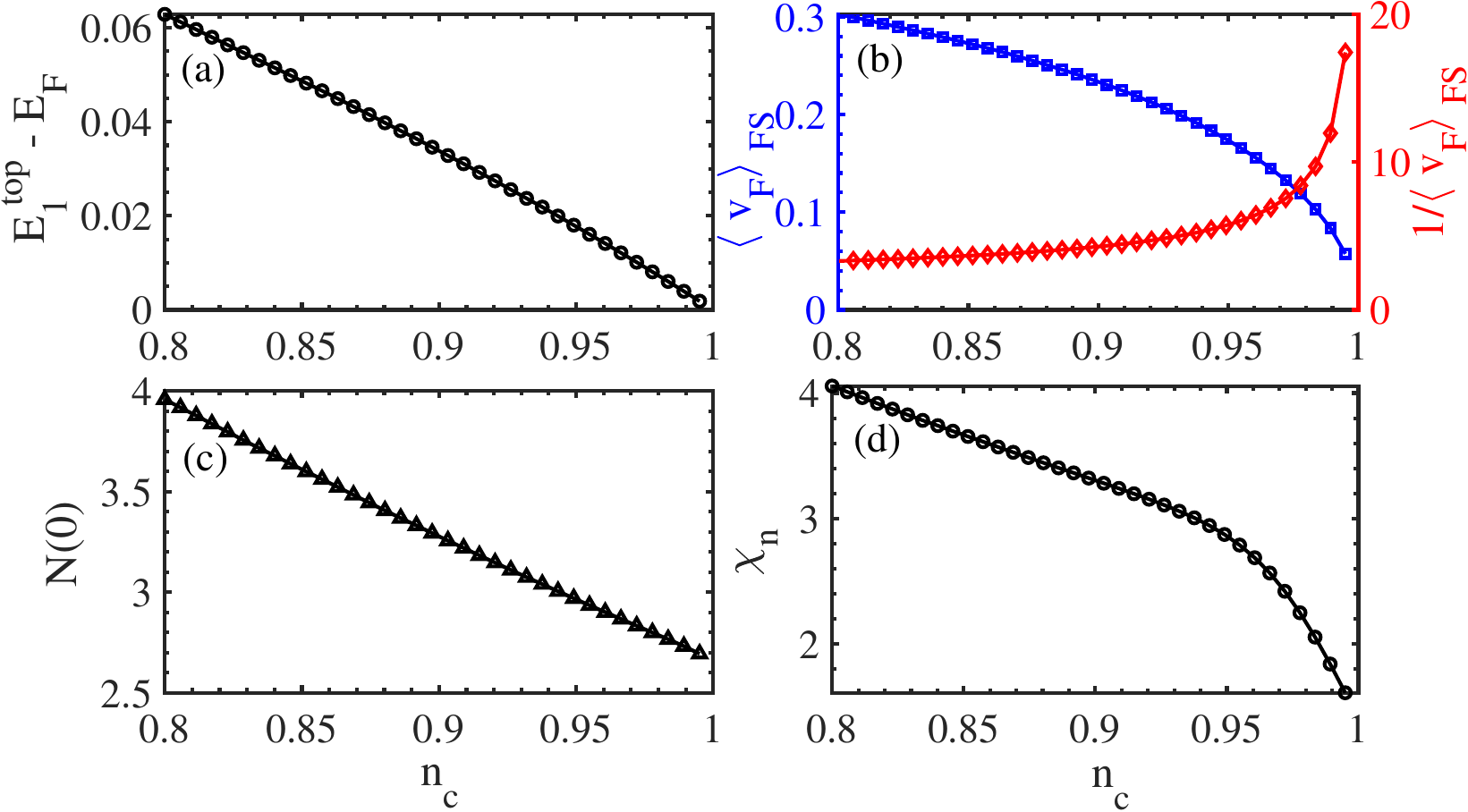}
  \caption{(Color online) Evolution of band-structure and Fermi-surface diagnostics as a function of the conduction filling $n_c$ at fixed $J_H/t=1$. (a) Energy separation between the Fermi level and the top of the lower hybridized band, $E_1^{\rm top}-E_F$. (b) Fermi-surface–averaged Fermi velocity $\langle v_F\rangle_{\rm FS}$ (blue) and its inverse $\langle 1/v_F\rangle_{\rm FS}$ (red). (c) Density of states at the Fermi level, $N(0)$. (d) Normal-state uniform spin susceptibility $\chi_n$ obtained from the mean-field solution. Together, these quantities demonstrate how increasing $n_c$ moves the Fermi level toward the weakly dispersive region at the top of the lower hybridized band.}
  \label{fig9}
\end{figure}

\begin{figure}
  \centering
  \includegraphics[width=0.5\textwidth]{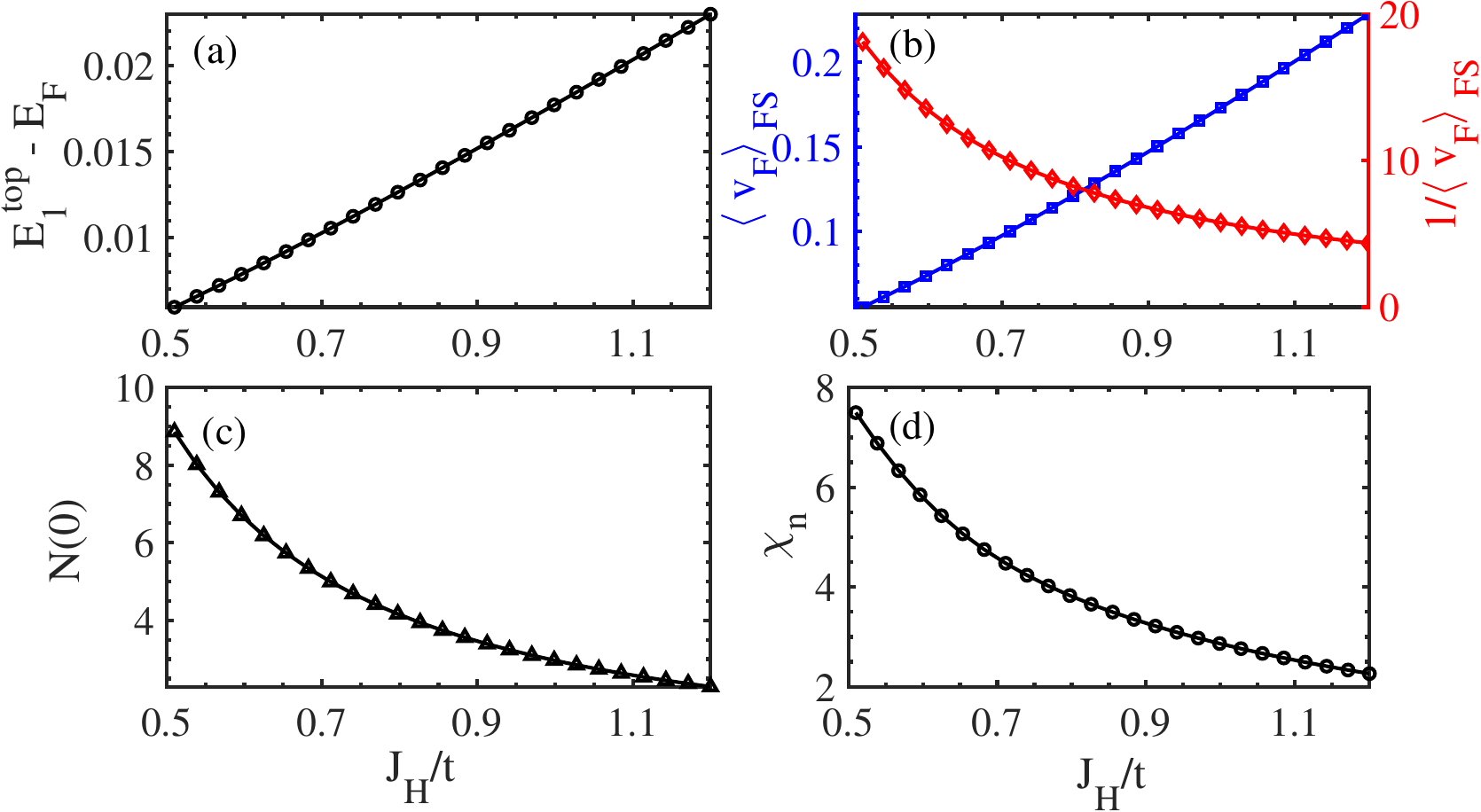}
  \caption{(Color online) Evolution of normal-state band and Fermi-surface properties as a function of $J_H/t$ at fixed filling $n_c=0.95$. (a) Energy separation $E_1^{\rm top}-E_F$. (b) Fermi-surface–averaged Fermi velocity $\langle v_F\rangle_{\rm FS}$ (blue) and its inverse $\langle 1/v_F\rangle_{\rm FS}$ (red). (c) Density of states at the Fermi level $N(0)$. (d) Normal-state spin susceptibility $\chi_n$.}
  \label{fig10}
\end{figure}

Fig.~\ref{fig8} shows that the ratio $r_{\mathrm{CC}}$ increases monotonically with $n_c$ throughout the $(J_H/t,n_c)$ plane, indicating a progressive enhancement of the Pauli-limited robustness as the Fermi level approaches the weakly dispersive (near-flat) region of the lower hybridized band.

%While Fig.~\ref{fig14} offers only a schematic view of how hybridization produces a weakly dispersive band top, the fully self-consistent dispersion in Fig.~\ref{fig11} verifies that the maximum of the lower hybridized band indeed forms a nearly flat region.  This establishes the microscopic structure underlying the monotonic growth of $r_{\mathrm{CC}}$ with $n_c$ in Fig.~\ref{fig8}, and sets the stage for the quantitative diagnostics shown in Fig.~\ref{fig9}.

To clarify the microscopic origin, we analyze the $J_H/t=1$ slice of Fig.~\ref{fig8} alongside the band and Fermi-surface diagnostics in Fig.~\ref{fig9}.  As $n_c$ increases, the Fermi level moves toward the weakly dispersive band top, as indicated by the steadily decreasing offset $E^{\mathrm{top}}_1 - E_F$ and the strong suppression of the Fermi-surface–averaged velocity.  These quantities therefore provide the most sensitive indicators of proximity to the near-flat region of the hybridized band.

Importantly, however, approaching this weakly dispersive regime does not lead to an enhanced density of states.  Both $N(0)$ and $\chi_n$ exhibit only a
mild decrease, not because the band is more dispersive, but because the Fermi surface shrinks rapidly as $n_c\!\to\!1$, reducing the available phase-space
length.  Thus, in the present parameter regime, Fermi-surface geometry rather than local band flatness governs the behavior of $N(0)$ and $\chi_n$ (see Appendix Fig.~\ref{fig12}).

Having identified this geometric–dynamical trend, we now address why proximity to the weakly dispersive region enhances the Pauli-limited field.  For this purpose, we analyze the zero-temperature free-energy balance between the superconducting and normal states within a minimal two-band description consisting of a light dispersive band $\xi_{1\boldsymbol{k}}$ and a weakly dispersive band $\xi_{2\boldsymbol{k}}\simeq -\mu_0$. Under a Zeeman field $h=\mu_{\mathrm B}B$, the Bogoliubov quasiparticle spectra take the form
\[
E_i(\boldsymbol{k})=\sqrt{\xi_{i\boldsymbol{k}}^2+\Delta_i^2},\qquad E_{i\sigma}(\boldsymbol{k})=E_i(\boldsymbol{k})-\sigma h .
\]
The free-energy difference $\Delta F(h)=F_S(h)-F_N(h)$ can be decomposed as
\[
\Delta F(h)=E_{\boldsymbol{cond}}^{(1)}+E_{\boldsymbol{cond}}^{(2)}  +\sum_i\!\left[L_i^{(S)}(h)-L_i^{(N)}(h)\right],
\]
where $L_{i}^{(N/S)}(h)$ denote the paramagnetic energies of band $i$ in the normal/superconducting states.  The finite-temperature expression
\[
L_i=-\frac{1}{\beta}\sum_{\boldsymbol{k}\sigma}\ln\!\left(1+e^{-\beta E_{i\boldsymbol{k}\sigma}}\right)
\]
reduces at $T\!\to\!0$ to contributions only from negative-energy quasiparticles:
\[
\lim_{\beta\to\infty} \!\left[-\frac{1}{\beta}\ln\!\left(1+e^{-\beta E_{i\boldsymbol{k}\sigma}}\right)\right]=E_{i\boldsymbol{k}\sigma}\,\Theta(-E_{i\boldsymbol{k}\sigma}).
\]

In the normal state ($\Delta_1=0$), the Zeeman field generates spin-flipped states in the window $0<|\xi_{1\boldsymbol{k}}|<h$, giving the standard Pauli
paramagnetic gain
\[
L_1^{(N)}(h)
= -\frac12 N_1(0)\,h^2 .
\]

In the superconducting state with $\Delta_1>0$, the spectrum satisfies $E_1(\boldsymbol{k})\ge\Delta_1$.  Thus for $h<\Delta_1$, both branches $E_{1\sigma}(\boldsymbol{k})$ remain positive and no negative-energy quasiparticles form:
\[
L_1^{(S)}(h)=0,\qquad (h<\Delta_1).
\]
When $h>\Delta_1$, states with $|\xi_{1\boldsymbol{k}}|<\sqrt{h^2-\Delta_1^2}$ acquire negative-energy branches $E_{1\uparrow}<0$, and the superconducting state begins to absorb paramagnetic energy:
\[
L_1^{(S)}(h)=N_1(0)\!\left[\frac{\Delta_1^2}{2}\ln\!\frac{h+\sqrt{h^2-\Delta_1^2}}{\Delta_1}-\frac{h}{2}\sqrt{h^2-\Delta_1^2}\right].
\]

For the weakly dispersive band $\xi_{2\boldsymbol{k}}\simeq -\mu_0$ with $\mu_0>h$, the energies $-\mu_0\mp h$ never cross the Fermi level, so
\[
L_2^{(N)}(h)\simeq 0 .
\]
Including pairing gives
\[
E_{2\sigma}(\boldsymbol{k})
=\sqrt{\mu_0^2+\Delta_2^2}-\sigma h,
\]
which remains positive for all $\boldsymbol{k}$ as long as $h\ll\sqrt{\mu_0^2+\Delta_2^2}$:
\[
L_2^{(S)}(h)\simeq 0 .
\]
Hence the weakly dispersive band produces negligible paramagnetic energy in either state, while its condensation energy $E_{\mathrm{cond}}^{(2)}\sim -\tfrac12 N_2(0)\Delta_2^2$ remains intact over a broad field interval.

Collecting all terms, the free-energy difference is
\[
\Delta F(h) =E_{\mathrm{cond}}^{(1)}+E_{\mathrm{cond}}^{(2)} +\bigl[L_1^{(S)}(h)-L_1^{(N)}(h)\bigr]
\]
because the weakly dispersive band satisfies $L_2^{(S)}(h)\simeq L_2^{(N)}(h)\simeq 0$.  Both $L_1^{(N)}(h)$ and $L_1^{(S)}(h)$ become more negative with increasing field, but the normal-state term grows much more rapidly due to the expanding spin-flip window $0<|\xi_1|<h$.   In contrast, the heavy band contributes almost no paramagnetic energy in either state, leaving its condensation-energy gain essentially unchanged.  As a result, the free-energy crossing $\Delta F(h_c)=0$ is pushed to fields far exceeding the conventional Pauli limit: superconductivity survives until the light band’s paramagnetic advantage finally overcomes the robust and nearly field-independent condensation energy provided by the weakly dispersive band.

This ``mismatch'' strongly diminishes the efficiency of Zeeman pair breaking and enhances the Pauli limit.  In our self-consistent calculations, $B_c$ is determined from $\Delta_{cf(f)}(B_c)\to0$; the extremely weak Zeeman response of near-flat quasiparticles allows $\Delta_{cf(f)}(B)$ to persist up to fields far exceeding $1.25\,T_c$, producing the large $r_{\mathrm{CC}}$ values in Fig.~\ref{fig8}.

At fixed filling \(n_c=0.95\), different pairing channels appear in distinct ranges of \(J_H\); for clarity we focus on the extended-\(s\) state, whose broad \(J_H\) window allows a clear view of the evolution of the diagnostic quantities.  As shown in Fig.~\ref{fig10}, increasing \(J_H\) pushes the Fermi level farther below the top of the lower hybridized band, yielding a monotonic increase of \(E^{\mathrm{top}}_1-E_F\) and driving the system away from the weakly dispersive region.  This shift is accompanied by a larger characteristic Fermi velocity—\(\langle v_F\rangle_{\mathrm{FS}}\) rises while \(1/\langle v_F\rangle_{\mathrm{FS}}\) decreases—and by concurrent reductions of the density of states \(N(0)\) and the Pauli susceptibility \(\chi_n\), consistent with the expected trend when moving away from a near-flat band.  Consequently, increasing \(J_H\) weakens the flat-band–assisted enhancement of the Pauli-limited field and reduces \(r_{\mathrm{CC}}\) at large \(J_H\).

Taken together, Figs.~\ref{fig9} and \ref{fig10} show that the filling \(n_c\) is the primary knob controlling proximity to the weakly dispersive part of the hybridized band, which naturally explains the robust increase of \(r_{\mathrm{CC}}\) with \(n_c\) in Fig.~\ref{fig8}.  In contrast, \(J_H\) mainly modulates hybridization and dispersion, generating pairing-channel–dependent variations of \(r_{\mathrm{CC}}\). 

\section{Conclusion}\label{sec:level5}

We have investigated the magnetic response of superconductivity in the two-dimensional Kondo--Heisenberg model using a self-consistent mean-field approach.  This minimal framework incorporates strong electronic correlations together with $c$--$f$ hybridization, and thus provides a controlled setting for examining unconventional behavior of the upper critical field.

By computing $T_c(B)$ and extracting the Clogston--Chandrasekhar ratio $r_{\mathrm{CC}}$, we find that its magnitude is highly sensitive to both the electronic filling and the exchange coupling.  In particular, $r_{\mathrm{CC}}$ increases significantly as $n_c$ approaches half filling.  Analysis of the band structure and Fermi-surface evolution shows that this trend is consistent with a ``near-flat-band'' mechanism: as the Fermi level moves into a weakly dispersive region of the hybridized band, the Fermi velocity is reduced and the normal-state spin-polarization energy cost decreases, thereby weakening Pauli pair breaking and enhancing the effective Pauli limiting field.

At fixed filling, the dependence of $r_{\mathrm{CC}}$ on $J_H$ is strongly channel selective.  As a representative example, for extended-$s$ pairing, increasing $J_H$ moves the Fermi level away from the near-flat region and concomitantly reduces $r_{\mathrm{CC}}$.  This highlights that the Pauli limiting field in correlated, multicomponent systems cannot be characterized solely by the zero-field gap magnitude but is also controlled by band dispersion and pairing structure.

Overall, our results show that a hybridized band approaching a flat-band regime can naturally generate an enhanced Pauli critical field even within a purely spin-singlet framework with an isotropic $g$ factor, without invoking additional mechanisms.  This provides a unified band-structure viewpoint for understanding apparent Pauli-limit violations in heavy-fermion and related superconductors, and establishes a useful baseline for future studies that incorporate orbital-selective pairing, spin--orbit coupling, or correlations beyond mean field~\cite{PhysRevB.108.214511,PhysRevB.89.024509,Nica2022Frontiers}.

\begin{acknowledgments}
This work was supported by the National Natural Science Foundation of China (Grant No.~12247101), the Fundamental Research Funds for the Central Universities (Grant No.~lzujbky-2024-jdzx06), the Natural Science Foundation of Gansu Province (Grant Nos.~22JR5RA389 and 25JRRA799), and the National ``111 Center''
 for Collaborative Research (Grant No.~B20063).
\end{acknowledgments}

\appendix
\section{Fermi-surface extraction}

\begin{figure}[t]
  \centering
  \includegraphics[width=0.48\textwidth]{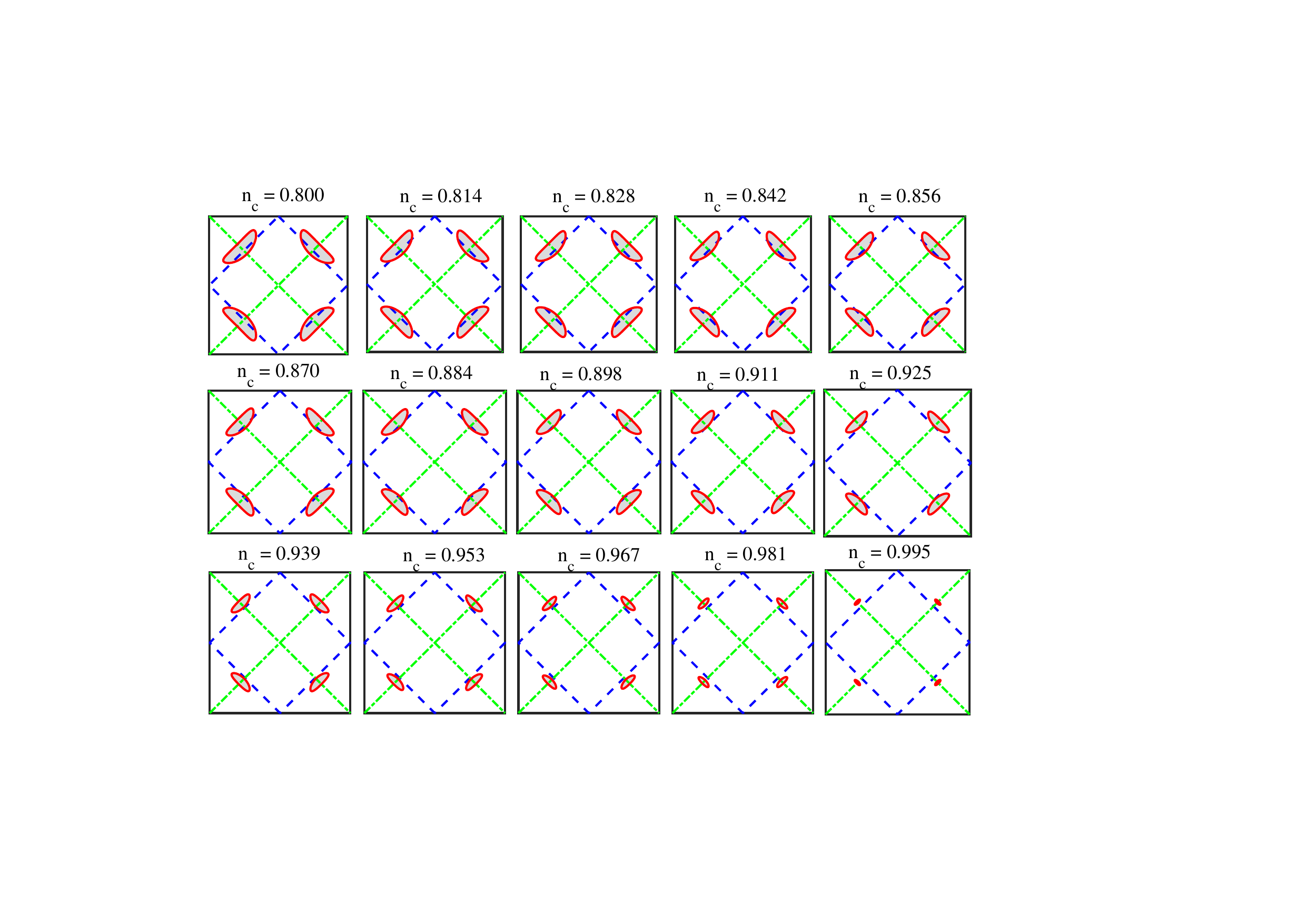}
  \vspace{3mm}
  \includegraphics[width=0.48\textwidth]{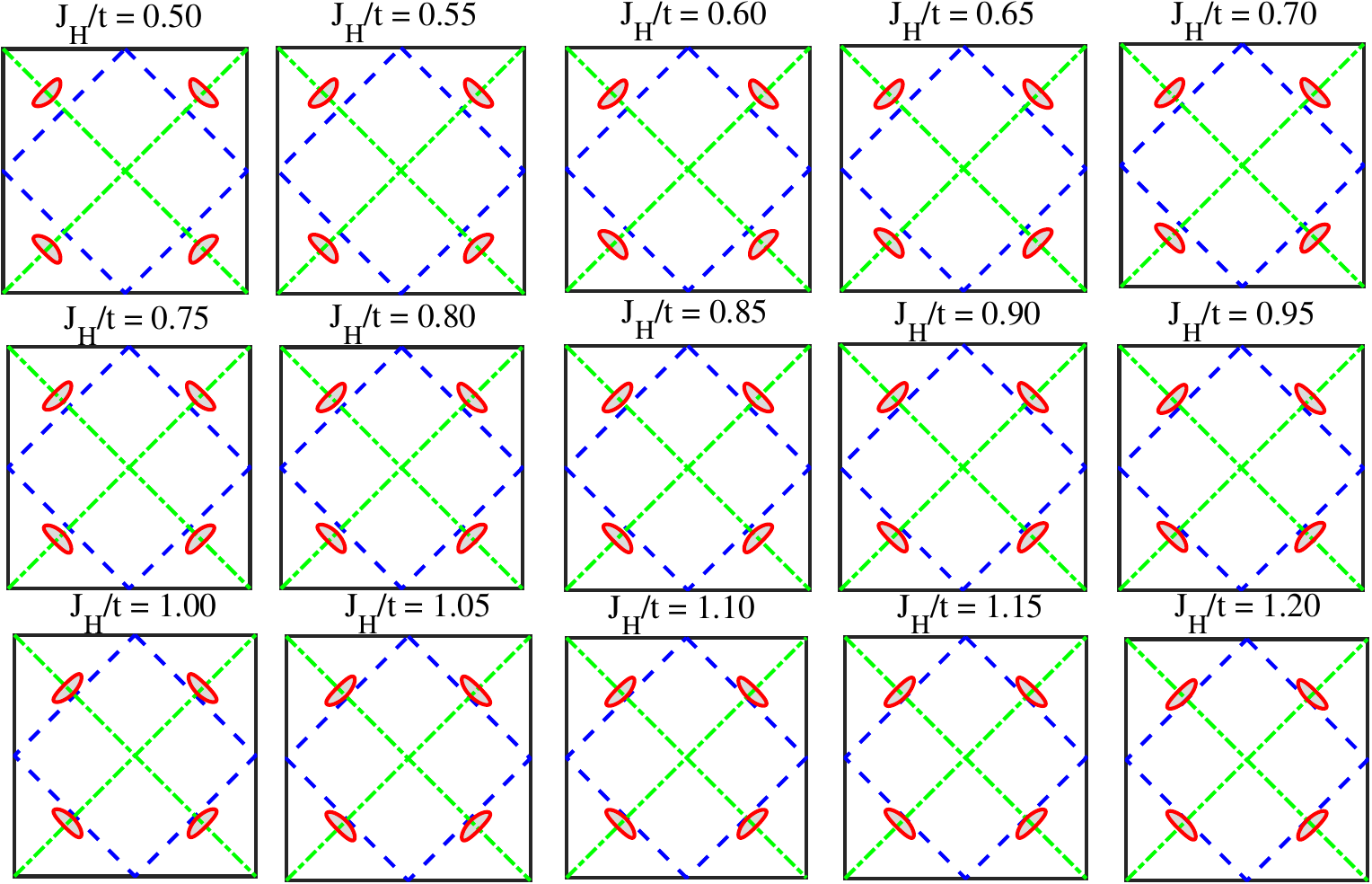}
  \caption{(Color online) (a) Evolution of the heavy-fermion Fermi surface with filling $n_c$ for $J_H/t=1$.  The red solid curves denote the Fermi surface of the lower hybridized band, while the green dash–dotted and blue dashed lines indicate the noninteracting Fermi surfaces of the $c$ electrons and $f$ spinons.
  (b) Evolution of the Fermi surface as a function of $J_H/t$ at fixed filling $n_c=0.95$.  Each panel corresponds to a distinct parameter value as labeled.}
  \label{fig12}
\end{figure}

This appendix summarizes the procedure used to generate the Fermi surfaces shown in Fig.~\ref{fig12}(a).  For a given set of parameters $(J_K/t,\,t'/t,\,J_H/t,\,n_c)$, we first solve the normal-state mean-field equations to obtain the self-consistent values of $(\chi,\,V,\,\lambda,\,\mu)$. These values are inserted into the two hybridized bands [Eq.~\ref{1}],
\begin{equation}
E^{\pm}_{\boldsymbol k}
= \frac12\!\left[(\epsilon_{\boldsymbol k}+\chi_{\boldsymbol k})\pm \sqrt{(\epsilon_{\boldsymbol k}-\chi_{\boldsymbol k})^{2}+ (J_K V)^2}\right].
\end{equation}

The heavy-fermion Fermi surface is obtained by solving $E^{-}_{\boldsymbol k}=0$ and numerically tracing sign changes of $E^{-}_{\boldsymbol k}$ on an $800\times800$ momentum grid using linear interpolation.  The resulting closed contours are displayed as the red solid curves in Fig.~\ref{fig12}(a).  For comparison, the green dash–dotted and blue dashed lines denote the bare (nonhybridized) Fermi surfaces of the $c$ electrons and $f$ spinons, respectively.

In Fig.~\ref{fig12}(a), the evolution of the heavy-fermion Fermi surface with increasing filling $n_c$ is clearly visible.  At low fillings, the Fermi surface is relatively large and resides in a strongly dispersive region of the lower hybridized band.  As $n_c$ increases, the Fermi surface shrinks continuously and moves toward the weakly dispersive band-top region identified in Fig.~\ref{fig9}.  Near half filling ($n_c\!\to\!1$), the Fermi surface becomes very small and is pinned close to this weakly dispersive band-top region.  This reflects the monotonic increase of the chemical potential, which pushes $E_F$ upward toward the low-curvature portion of $E_1(\boldsymbol k)$ and thereby enhances the heavy-fermion character through a suppressed Fermi velocity.

In contrast to the strong filling dependence in Fig.~\ref{fig12}(a), the Fermi surface is only weakly affected by variations of $J_H/t$ at fixed $n_c=0.95$, as illustrated in Fig.~\ref{fig12}(b).  Although $J_H$ modifies the hybridization strength and the detailed band curvature, the overall shape and size of the heavy-fermion pockets remain nearly unchanged.  This indicates that the approach of $E_F$ toward the weakly dispersive band-top region is governed primarily by the filling rather than by the exchange coupling.

\bibliographystyle{apsrev4-2} 
\bibliography{Refs}
\end{document}